\documentclass[12pt]{article}
\usepackage{amsmath}
\usepackage{graphicx}
\usepackage{enumerate}
\usepackage{natbib}
\usepackage{url} 

\newcommand{\blind}{1}

\usepackage{mathtools}
\usepackage{amsfonts}
\usepackage{algorithm}      
\usepackage[noend]{algpseudocode} 
\usepackage{wrapfig}
\usepackage{caption}
\usepackage{subcaption}
\usepackage{xcolor}         
\RequirePackage[colorlinks,citecolor=blue,urlcolor=blue]{hyperref}

\newcommand{\BlackBox}{\rule{1.5ex}{1.5ex}}  

\newtheorem{theorem}{Theorem}

\newtheorem{assume}{Assumption}
 
\newtheorem{proposition}{Proposition} 
\newtheorem{remark}{Remark}

\usepackage{mathrsfs}
\usepackage{multirow}

\begin{document}

\def\spacingset#1{\renewcommand{\baselinestretch}%
{#1}\small\normalsize} \spacingset{1}


\let\oldemptyset\emptyset
\let\emptyset\varnothing

\def\U{\mathbf{U}}
\def\S{\mathbf{S}}
\def\V{\mathbf{V}}
\def\W{\mathbf{W}}
\def\X{\mathbf{X}}
\def\Y{\mathbf{Y}}
\def\Z{\mathbf{Z}}
\def\A{\mathbf{A}}
\def\a{\mathbf{a}}
\def\Ssc{\mathcal{S}}
\def\Tsc{\mathcal{T}}
\def\Bsc{\mathcal{B}}
\def\e{\mathbf{e}}
\def\c{\mathbf{c}}
\def\u{\mathbf{u}}
\def\w{\mathbf{w}}
\def\x{\mathbf{x}}
\def\y{\mathbf{y}}
\def\z{\mathbf{z}}
\def\H{\mathbf{H}}

\def\Dscr{\mathscr{D}}

\def\bSigma{\boldsymbol{\Sigma}}
\def\bPhi{\boldsymbol{\Phi}}
\def\bbeta{\boldsymbol{\beta}}
\def\balpha{\boldsymbol{\alpha}}
\def\bgamma{\boldsymbol{\gamma}}
\def\btheta{\boldsymbol{\theta}}
\def\bpsi{\boldsymbol{\psi}}
\def\bphi{\boldsymbol{\phi}}
\def\bzero{\mathbf{0}}
\def\AUC{\mathbf{AUC}} 
\def\TPR{\mathbf{TPR}} 
\def\FPR{\mathbf{FPR}}
\def\TP{\mathbf{TP}} 
\def\FP{\mathbf{FP}}

\def\PPV{\mathbf{PPV}} 
\def\NPR{\mathbf{NPV}} 
\def\ROC{\mathbf{ROC}} 
\def\Pbb{\mathbb{P}}
\def\Psc{\mathscr{P}}
\def\Pschat{\widehat\Psc}
\def\roc{{\sf roc}}
\def\cS{\mathcal{S}}
\def\cT{\mathcal{T}}
\def\cI{\mathcal{I}}
\def\Msc{\mathcal{M}}
\def\Csc{\mathcal{C}}
\def \Wsc{\mathcal{W}}
\def\subcal{_{\sf cal}}
\def\bepsilon{\mathbf{\epsilon}}
\def\Qcal{\mathcal{Q}}
\def\Bsc{\mathcal{B}}
\def\Esc{\mathcal{E}}
\def\Ksc{\mathcal{K}}
\def\D{\mathbf{D}}
\def\argmindum{\mathop{\mbox{argmin}}}
\def\argmin#1{\argmindum_{#1}}
\def\argmaxdum{\mathop{\mbox{argmax}}}
\def\argmax#1{\argmaxdum_{#1}}

\def \P {\mathbb{P}}
\def \E {\mathbb{E}}
\def \R {\mathbb{R}}
\def \I {\mathbb{I}}
\def\trans{^{\scriptscriptstyle \sf T}}
\def\b{\mathbf{b}}

\def\Cupsub{{\scriptscriptstyle \cup}}
\def\Tscsub{{\scriptscriptstyle \Tsc}}
\def\Sscsub{{\scriptscriptstyle \Ssc}}
\def\Kscsub{{\scriptscriptstyle \Ksc}}

\def\suphalf{^{\frac{1}{2}}}

\def\bdelta{\boldsymbol{\delta}}
\def\Fsc{\mathcal{F}}
\def\Gsc{\mathcal{G}}
\def\Lsc{\mathcal{L}}
\def\Qsc{\mathcal{Q}}
\def\Hsc{\mathcal{H}}
\def\u{\boldsymbol{u}}

\def\subTsc{_{\scriptscriptstyle \Tsc}}
\def\subSsc{_{\scriptscriptstyle \Ssc}}
\def\Pbb{\mathbb{P}}

\if1\blind
{
  \title{\bf Doubly Robust Augmented Model Accuracy Transfer Inference with High Dimensional Features}
  \author{ Doudou Zhou$^{1*}$, Molei Liu$^{2}$\footnote{Zhou and Liu contributed equally.}, Mengyan Li$^{3}$, Tianxi Cai$^{4,5}$ \bigskip \\
\small 
$^1${Department of Statistics, University of California, Davis} \\
\small 
$^2${Department of Biostatistics, Columbia University Mailman School of Public Health} \\
\small 
$^3${Department of Mathematical Sciences, Bentley University} \\
\small
$^4${Department of Biostatistics, Harvard T.H. Chan School of Public Health}\\
\small
$^5${Department of Biomedical Informatics, Harvard Medical School}
}
  \maketitle
} \fi

\if0\blind
{
  \bigskip
  \bigskip
  \bigskip
  \begin{center}
    {\LARGE\bf Doubly Robust Augmented Model Accuracy Transfer Inference with High Dimensional Features}
\end{center}
  \medskip
} \fi

\bigskip
\begin{abstract}
Due to label scarcity and covariate shift occurring frequently in real-world studies, transfer learning has become an essential technique to train models generalizable to some target populations using existing labeled source data. Most existing transfer learning research has been focused on model estimation, while there is a paucity of literature on transfer inference for model accuracy despite its importance. We propose a novel {\bf D}oubly {\bf R}obust {\bf A}ugmented {\bf M}odel {\bf A}ccuracy {\bf T}ransfer {\bf I}nferen{\bf C}e (DRAMATIC) method for point and interval estimation of commonly used classification performance measures in an unlabeled target population using labeled source data. Specifically, DRAMATIC derives and evaluates the risk model for a binary response $Y$ against some low dimensional predictors $\A$ on the target population, leveraging $Y$ from source data only and high-dimensional adjustment features $\X$ from both the source and target data. The method builds on top of an imputation model for the mean of $Y\mid \X$ and a density ratio model characterizing the distributional shift of $\X$ between the source and target populations. We first develop a doubly robust estimating equation on the two models for the risk model of $Y\mid \A$ on the target population. 
Then a doubly robust estimation is constructed for the receiver operating characteristic (ROC) curve used for the model accuracy evaluation. During the process, we mitigate the regularization biases incurred by the high-dimensional nuisance estimators of the imputation and density ratio models by constructing appropriate calibrated moment equations. The proposed estimators are doubly robust in the sense that they are $n\suphalf$ consistent when at least one model is correctly specified, and certain model sparsity assumptions hold. Simulation results demonstrate that the point estimation has negligible bias and the confidence intervals derived by DRAMATIC attain satisfactory empirical coverage levels. We further illustrate the utility of our method to transfer the genetic risk prediction model, and its accuracy evaluation for type II diabetes across two patient cohorts in Mass General Brigham (MGB) collected using different sampling mechanisms and at different time points.
\end{abstract}

\noindent%
\vfill

\newpage
\spacingset{1.9} 
\section{Introduction}

\subsection{Background}

While cohort studies remain to be critical sources for studying disease progression and treatment response, they have limitations, including the generalizability of the study findings to the real world, the limited ability to test broader hypotheses, and the cost associated with performing these studies. In recent years, due to the increasing adoption of electronic health records (EHR) and the linkage of EHR with specimen bio-repositories, integrated large datasets now exist as a new data source for deriving real-world predictive models. For example, the Massachusetts General and Brigham (MGB) Healthcare Biobank \citep{castro2022mass} and the million veteran's project \citep{gaziano2016million} contain a wealth of both clinical and genomic data for patients enrolled in the biobanks. These integrated datasets open opportunities for developing accurate EHR-based personalized risk prediction models, which can be easily integrated into clinical workflow to assist clinical decision-making. These models can also be contrasted and integrated with models derived from traditional cohort studies to improve generalizability. 

Efficiently deriving and evaluating prediction models using EHR data remains challenging due to practical and methodological obstacles. 
First, gold standard labels on disease outcomes are typically not readily available in EHR and require accessibility of clinical notes and labor-intensive manual annotations. For example, at MGB, only $\sim 74\%$ of patients with at least one diagnostic code of type II diabetes (T2D) have T2D \citep{liao2019high}. On the other hand, verifying true T2D status via manual chart review is laborious and thus only available for a small number of patients at MGB. Existing methods for deriving EHR risk models typically fit prediction models to inaccurate proxies of the true outcome, such as the presence or absence of a billing code or a classification derived from a machine learning algorithm \citep{kurreeman2011genetic}. The misclassification error in the proxies can lead to biases in the derived prediction model and the estimated accuracy of the trained model. 

Another challenge associated with ensuring the generalizability of EHR derived model arises from the heterogeneity between different EHR cohorts. It is well-known that patient populations may differ substantially across different healthcare systems, leading to significant heterogeneity in the performance of prediction models \citep{rasmy2018study}. Risk prediction models derived to optimize one EHR cohort may not perform well for another cohort. In addition, data for EHR studies are often refreshed over time as new patients enter the study, and the data of existing patients get updated. Thus, EHR data from two different time points of the same institution can have a significant shift in their distribution. 

Both the scarcity of gold standard labels on disease outcomes and the covariate shift poses great challenges to generalizing knowledge about EHR-derived risk models from a labeled source population to a target population where gold standard labels are typically not available. To transfer such knowledge, much recent literature has been focused on transfer learning techniques for model estimation \citep{liu2020doubly,geva2021high}, such as the transportability of prediction algorithms trained from EHR data \citep{weng2020deep}. On the other hand, there is a paucity of literature on transfer learning of model performance. In this paper, we propose a novel {\bf D}oubly {\bf R}obust {\bf A}ugmented {\bf M}odel {\bf A}ccuracy {\bf T}ransfer {\bf I}nferen{\bf C}e (DRAMATIC) procedure to enable point and interval estimation for commonly used classification performance measures in an unlabeled target population using labeled source data. Specifically, the proposed DRAMATIC procedure develops and evaluates a prediction model for a binary response $Y$ given low dimensional predictors $\A$ in the target population by leveraging source data on $Y$ and high-dimensional features $\X = (\A\trans, \W\trans)\trans$ and only requires to be shared between the source and target data. We next detail the problem of interest and highlight our contributions in view of existing literature. 

\subsection{Problem statement}
Let $Y$ denote the binary outcome of interest and $\A_{q\times 1} = (A_1,\dots,A_q)\trans$ denote a vector of predictors used to predict $Y$ and the first element of $\A$ being 1. Our goal is to develop and evaluate a risk model predicting $Y$ with $\A$ for a target population $\Tsc$ using observed data on $n$ samples of $(Y, \X\trans)\trans$ in a source population $\Ssc$ and $N$ samples of $\X$ in $\Tsc$, where $\X=(\A\trans,\W\trans)\trans$ and $\W_{p\times 1} =  (W_1,\dots,W_p)\trans$ denotes confounding (adjustment) covariates that may relate to $Y$ and/or the distributional shift between $\Tsc$ and $\Ssc$. We use $S=1$ to indicate that the sample is labeled from $\Ssc$ and $S = 0$ for the unlabelled target data. We assume that $q$, the dimension of risk factors, is fixed but allow $p$, the dimension of the covariates $\W$ needed to enable transfer learning, to be high, i.e., $p \gg \min(n,N)$. In an EHR-driven genetic risk prediction study, examples of $\A$ include relevant genetic markers and demographic variables, while $\W$ may include a large number of EHR proxies of $Y$, including relevant diagnostic codes, medication prescriptions, laboratory tests, as well as mentions of clinical terms extracted from free text clinical notes via natural language processing. Hence $Y \mid \A$ may differ between the target and source populations, but $Y \mid \A,\W$ is assumed to be the same.

To develop the risk model for $Y \mid \A$ in $\Tsc$, we assume a {\em working} logistic regression model 
$$
\P\subTsc(Y = 1 \mid \A) = \P(Y = 1 \mid \A,S=0)  = g(\A\trans\bbeta) \quad \mbox{with}\quad g(x)=1/(1+e^{-x}),
$$
which is not required to hold throughout. Here we define $\P\subTsc(\cdot) = \P ( \cdot \mid S=0)$, $\P\subSsc(\cdot) = \P ( \cdot \mid S=1)$, $\E\subTsc[\cdot] = \E[ \cdot \mid S=0]$ and $\E\subSsc[\cdot] = \E[ \cdot \mid S=1]$. 
Let the observed data be $\Dscr = \{\D_i =(S_i Y_i, \X_i\trans , S_i)\trans: i = 1, 2, \dots, n+N \}$ where we let $S_i=\I(1 \le i \le n)$ without loss of generality and  $\I(\cdot)$ is the indicator function. Assume that the samples are generated following $
Y,\X\mid S=s\sim  p_{\X|S=s}(\x)\cdot p_{Y|\X}(y),
$
where $p_{\X|S=s}(\cdot)$ and $p_{Y|\X}(\cdot)$ represent the probability mass/density function of $\X$ given $S$ and $Y$ given $\X$ respectively. Here, our underlying assumption is that distribution of covariates $\X$ is different across $\Ssc$ and $\Tsc$ while the conditional distribution of the response $Y$ given $\X$ is the same across the two populations. We define the density ratio function as $h_0(\x) = p_{\X|S=0}(\x)/p_{\X|S=1}(\x)$ and the conditional mean of $Y$ on the two populations as $ r_0(\x)=\E[Y\mid\X=\x]$.

Let $\bbeta_0$ be the outcome parameter of our interests that solves the estimating equation: 
\[
\E\subTsc [\A\{Y-g(\A\trans \bbeta)\}] = \bzero \,.
\]
To evaluate the accuracy of the risk model $g(\A\trans\bbeta_0)$ on $\Tsc$, we aim at inferring the commonly used receiver operating characteristic (ROC) curve of $Y$ against $\A\trans\bbeta_0$. 
Specifically, for any given cutoff value $c$, we may classify a patient as high risk if $\A\trans\bbeta_0\ge c$, and low risk otherwise. Then the true positive rate (TPR) and false positive rate (FPR) associated with the classification rule $\I(\A\trans\bbeta_0 \ge c)$ on $\Tsc$ are defined as
\[
\TPR(c) = \Pbb\subTsc(\A\trans\bbeta_0 \geq c \mid Y=1) \text{ and } \FPR(c) = \Pbb\subTsc(\A\trans\bbeta_0\geq c \mid Y=0).
\]
Then the trade-off between TPR and FPR can be characterized by the ROC curve: $\ROC(u)=\TPR\{\FPR^{-1}(u)\}$ \citep{gerds2008performance}. The overall classification performance of the risk model can be summarized based on the area under the ROC curve (AUC): $\AUC=\int_0^1\ROC(u) {\rm d} u$. Our goal is to estimate the model parameter $\bbeta_0$ and the accuracy measures $\ROC(\cdot)$ and $\AUC$ on $\Tsc$ using data $\Dscr$, in which $Y$ is only observed for subjects in $\Ssc$ and the covariates $\X$ characterizing the shift between $\Ssc$ and $\Tsc$ are of high or ultra-high-dimensionality. 

\subsection{Related literature}

Transfer learning aims at using the information from the source population to aid the task in the target population \citep{torrey2010transfer}, with diverse 
real-world applications  \citep[e.g.]{ liu2006value,zhu2011heterogeneous,wiens2014study}. Existing machine learning methods have mainly focused on algorithms with little discussion about the theoretical guarantee, 
while theoretically justified transfer learning methods have largely focused on the estimation of low dimensional regression models. For example, \cite{huang2007correcting} proposed an importance weighting procedure based on kernel mean matching to correct for covariate shift in regression, \cite{reddi2015doubly} proposed a robust transfer learning strategy for removing bias while retaining small variance in parametric regression models, and \cite{liu2020doubly} introduced an imputation model for the response to augment importance weighting and achieved model double robustness for M-estimation. 

In the era of big data, more and more application fields like EHR studies involve a potentially high-dimensional set of features in knowledge transferring to characterize the distributional shift between the source and target populations. This poses additional methodological challenges to transfer learning due to the inherent biases induced by the regularization needed to overcome the high dimensionality. Transfer learning for high-dimensional regression models has been an active area of research in recent years \citep[e.g.]{li2020transfer,he2021transfer, bastani2021predicting,tian2021transfer,li2022transfer}. However, limited literature exists for transfer learning of prediction performance measures in high-dimensional settings. 

Existing transfer learning methods for model prediction performance can only handle the distributional shift of low-dimensional features. For example, \cite{rotnitzky2006doubly} developed estimators of the AUC of a single diagnostic test that adjust for selection verification. Their estimator is doubly robust in the sense of being consistent and asymptotically normal if either the outcome model or the selection-into-verification model is correct.  \cite{liu2010model} proposed a likelihood approach to estimate the nonignorable parameter, which is not identifiable in the previous works, and proved the asymptotic normality for their AUC estimators. Semiparametric estimation of the covariate specific ROC curves with a partial missing gold standard was developed by \cite{liu2011semiparametric}. Similarly, \cite{long2011robust} aimed at estimating the AUC of a disease status given a biomarker value under a setting where the biomarker values are missing for some subjects while disease status is always confirmed and a set of auxiliary variables are fully observed. 

The aforementioned methods focused on evaluating the prediction accuracy of a single predictor with low-dimensional adjustment variables for covariate shift. To the best of our knowledge, no existing work considers transfer inference for prediction accuracy of a fitted regression model in the presence of high-dimensional variables for covariates shift, which are substantially challenging due to the excessive (first-order) bias induced by high-dimensional nuisance estimators. Consequently, special debiasing techniques are needed to overcome biases induced by the high-dimensional nuisance models. 

The recent works of doubly robust estimation with high-dimensional nuisance models are technically relevant to the calibrated nuisance model estimation step in our DRAMATIC method  \citep{tan2020model,ning2020robust,dukes2020inference,liu2020double}. For example, \cite{tan2020model} considered the problem of estimating average treatment effects (ATE) with high-dimensional confounding covariates under certain sparse conditions. They proposed using the calibrated estimation equations to reduce the bias incurred by the regularized error of high-dimensional regression. Their estimators are doubly robust, which remains consistent if the propensity score model or the outcome regression model is correctly specified. The calibration idea has also been adopted for ATE estimation in \cite{ning2020robust}  via covariate balancing and for estimating the parametric components in a semiparametric model \citep{dukes2020inference}. \cite{liu2020double} estimated the parametric component of a logistic partially linear model by deriving certain moment equations to calibrate the first-order bias of the nuisance models. Further extensions have been recently studied, for example, to alleviate the ultra-sparsity assumptions of the nuisance models at the expense of requiring both models to be correct \citep{bradic2019sparsity}; and to solve general semiparametric estimating equations 
\citep{smucler2019unifying,ghosh2020doubly}. However, none of the existing methods can be easily applied to estimate a (working) logistic model and its associated ROC curve. Estimation for the ROC curve is particularly challenging since the estimator needs to converge uniformly for all points on the curve, which necessitates controlling an infinite number of moment equations simultaneously.

\subsection{Our contribution}
Our proposed DRAMATIC method is an efficient transfer learning procedure for estimating the model coefficients and accuracy parameters that overcomes the high-dimensionality of the nuisance models while attaining robustness to model misspecifications. It involves a calibration procedure to estimate the targeted logistic regression coefficients and the TPR, FPR, ROC, and AUC parameters and attains the desirable $n\suphalf$-consistency if either the high-dimensional imputation model or the high-dimensional density ratio model is correctly specified. 

Our work is the first one to consider a doubly robust estimation of the ROC curve of a working classification model under the high-dimensional covariate shift regime. In this process, the excessive regularization bias incurred by the high-dimensional nuisance estimators is removed by constructing a proper set of calibrated moment equations accommodating all cutoff points on the ROC curve. In specific, we propose a ``nearest quantile" strategy that ensures computation efficiency and the uniform $n\suphalf$-convergence of the ROC estimators by avoiding the demand of fitting the calibrated nuisance models on the extreme left side of the ROC curve, which would result in excessive estimation errors due to the poor effective sample size. Thus, our method also exhibits novelty in view of calibrating high-dimensional nuisance models for semiparametric estimation because non-existing methods in this field can be used to calibrate the doubly robust estimators of a curve with infinite points and achieve simultaneous convergence.



\section{DRAMATIC Procedure}
\label{sec:method}

\subsection{Outline and technical challenges}\label{sec:method:dr:cons}
Let $\Pschat\subSsc=n^{-1} \sum_{i=1}^n$, $\Pschat\subTsc=N^{-1} \sum_{i=n+1}^{N+n}$ and $\Pschat_{\Sscsub\Cupsub\Tscsub}={(N+n)}^{-1} \sum_{i=1}^{N+n}$ represent the empirical mean operators for the source samples, target samples and their union, respectively. Two common approaches to tackle the problem are (i) importance weighting and (ii) imputation. The former uses an estimated density ratio function $\widehat h(\cdot)$ to weight the source data which adjusts for covariate shift by solving the estimation equation
\begin{equation}
\Pschat\subSsc\widehat{h}(\X_i) \A_i \{ Y_i - g(\A_i\trans \bbeta) \} = \bzero\,,
\label{eq1}
\end{equation}
and the latter imputes the outcomes of the target data using an imputation function $\widehat r (\cdot)$ fitted by the source data, then solves the estimation equation
\begin{equation}
\Pschat\subTsc \A_i \{ \widehat r (\X_i) - g(\A_i\trans \bbeta) \} = \bzero \,.
\label{eq2}
\end{equation}
Although the two approaches are simple and natural, they are prone to excessive bias due to the potential model misspecification and the estimation errors of the nuisance parameters. Specifically, the importance weighting method leads to an inconsistent solution when the density ratio model $h_0$ is misspecified, and the imputation method is invalid when the imputation model $r_0$ is misspecified. 

We overcome these challenges by augmenting \eqref{eq1} and \eqref{eq2} to yield a consistent estimator of $\bbeta_0$ when either the density ratio model or the imputation model is correctly specified. To this end, we first specify the two high-dimensional nuisance models in parametric forms: $h(\x) = \exp(\x\trans\balpha)$ and $r(\x)=g(\x\trans\bgamma)$, which are not required to hold simultaneously. Let $\widehat\balpha$ and $\widehat\bgamma$ be some estimators for the nuisance parameters $\balpha$ and $\bgamma$, obtained using $\ell_1$-regularized regression. We then introduce the doubly robust estimator for $\bbeta$, denoted by $\widehat\bbeta$, as the solution to
\begin{equation}
   \Esc(\bbeta, \widehat\balpha, \widehat\bgamma ) := \Pschat\subSsc \exp(\X_i\trans\widehat\balpha) \A_i \{  Y_i - g(\X_i\trans\widehat\bgamma)\} + \Pschat\subTsc \A_i \{ g(\X_i\trans\widehat\bgamma) - g(\A_i\trans \bbeta) \} = \bzero \,.
\label{aeq}    
\end{equation}
The double robustness of the estimation equation \eqref{aeq} is validated in Proposition \ref{lem:double}. 
\begin{proposition}
For $\balpha_0 \in \R^p$ and $\bgamma_0 \in \R^p$, let $\bar \bbeta$ be the solution of $\E \Esc(\bbeta, \balpha_0, \bgamma_0) = \bzero$. If $\balpha_0$ satisfies $h_0(\x) = \exp(\x\trans\balpha_0)$ or $\bgamma_0$ satisfies $r_0(\x)=g(\x\trans\bgamma_0)$, then $\bar \bbeta = \bbeta_0$. 
\label{lem:double}
\end{proposition} 
However, unlike the low dimensional setting, due to both the potential model misspecification and the excessive estimation errors of the $\ell_1$-regularized nuisance estimators, simply plugging in arbitrary estimators $\widehat\balpha$ and $\widehat\bbeta$ in \eqref{aeq} could result in excessive first order bias and thus estimators converging slower than the desirable parametric rate even when one nuisance model is correct. Inspired by \cite{tan2020model}, we propose a bias calibration method detailed in Section \ref{sec:method:cal:out} to handle this problem. 

For the accuracy parameters $\TPR(c)$ and $\FPR(c)$, we propose doubly robust estimators $\widehat \TPR(c)$ and $\widehat \FPR(c)$ through augmentation similar to the construction in \eqref{aeq}. Specifically, for any given cutoff value $c$, we construct the doubly robust estimators for the proportion of the true positives $\P(\A_i\trans\bbeta_0 \ge c, Y_i=1)$ and the proportion of false positives $\P(\A_i\trans\bbeta_0 \ge c, Y_i = 0)$  as: 
\begin{align*}
& \widehat \TP(c; \widehat \bbeta,  \widehat\balpha_c, \widehat\bgamma_c) = \Pschat\subSsc\I( \A_i\trans \widehat \bbeta \geq c) \exp(\X_i\trans \widehat \balpha_c) \{Y_i-g(\X_i\trans\widehat\bgamma_c)\} + \Pschat\subTsc \I(\A_i\trans \widehat \bbeta\geq c) g(\X_i\trans\widehat\bgamma_c) \,, \\
& \widehat \FP(c; \widehat \bbeta, \widehat\balpha_c, \widehat\bgamma_c) = \Pschat\subSsc \I(\A_i\trans \widehat \bbeta \geq c) \exp(\X_i\trans\widehat\balpha) \{g(\X_i\trans\widehat\bgamma_c) - Y_i\} + \Pschat\subTsc\I(\A_i\trans \widehat \bbeta \geq c) \{1-g(\X_i\trans\widehat\bgamma_c)\} \,,
\end{align*}
where the nuisance parameters $\widehat\balpha_c$ and  $\widehat\bgamma_c$ are dependent on $c$ as will be shown later. Then we take the TPR and FPR estimators as
\begin{align}
& \widehat \TPR(c) := \widehat \TPR(c; \widehat \bbeta, \widehat\balpha_c, \widehat\bgamma_c, \widehat\balpha_{-\infty}, \widehat\bgamma_{-\infty}) = \frac{\widehat \TP(c; \widehat \bbeta, \widehat\balpha_c, \widehat\bgamma_c)}{\widehat \TP(-\infty; \widehat \bbeta, \widehat\balpha_{-\infty}, \widehat\bgamma_{-\infty})} \,, \label{equ:roc:dr0} \\
& \widehat \FPR(c) := \widehat \FPR(c; \widehat \bbeta, \widehat\balpha_c, \widehat\bgamma_c, \widehat\balpha_{-\infty}, \widehat\bgamma_{-\infty}) = \frac{\widehat \FP(c; \widehat \bbeta, \widehat\balpha_c, \widehat\bgamma_c)}{\widehat \FP(-\infty; \widehat \bbeta,  \widehat\balpha_{-\infty}, \widehat\bgamma_{-\infty})} \,.
\label{equ:roc:dr}
\end{align}
Here, for different cut-off $c$ (and $-\infty$), we plug-in different estimators for the nuisance parameters $\balpha_0$ and $\bgamma_0$ in the doubly robust estimators of TPR and FPR since the calibration is specific to $c$. 
The estimators for ROC and AUC can be constructed directly using the TPR and FPR estimators in \eqref{equ:roc:dr0} and  \eqref{equ:roc:dr}, and maintain a similar doubly robust property. 

\begin{remark}
Constructing calibrated estimators $\widehat\balpha_c$ and $\widehat\bgamma_c$ is key to achieving doubly robust properties for the accuracy parameters in the high-dimensional setting since 
the nuisance estimators' excessive bias can prevent them from achieving the parametric rate. We detail in Section \ref{sec:cal:roc} the moment conditions needed to constraint the nuisance estimators to overcome the bias. Eliminating bias for the ROC estimators is substantially more challenging than that of $\bbeta_0$ since the curve involves an infinite number of points, leading to both statistical and computational challenges. Specifically, the effective sample size for a large cutoff value $c$ can be much smaller than $n$, i.e., $\Pschat\subSsc\I( \A_i\trans \widehat \bbeta \geq c)$ can be small,  which deteriorates the convergence rate of $\widehat \TPR(c)$ and $\widehat \FPR(c)$. 
\end{remark}

\begin{remark}
Carrying out the calibration step for all observed unique values of $\A_i\trans  \widehat\bbeta$ is computationally prohibitive when $n$ is not small. To solve these problems, we propose a novel ``nearest" quantile method that ensures a desirable convergence rate and requires fitting an acceptable number of calibrating equations. 
\end{remark}

We outline the DRAMATIC method in Algorithm \ref{alg} and detail the key steps in Sections \ref{sec:method:cal:out} and \ref{sec:cal:roc}. A multiplier bootstrap method for inference is described and discussed in Supplementary S2, 
whose validity is ensured by the asymptotic normality of the proposed estimators as established in Section \ref{sec:thm}.

\begin{algorithm}
    \caption{Outline of DRAMATIC.}  \label{alg}
    \textbf{Input:} $\Dscr = \{\D_i =(S_i Y_i, \X_i\trans , S_i)\trans:  i = 1, 2, \dots, n+N \}$;  
    
    Constructing DR estimator $\widehat \bbeta$ of $\bbeta_0$:
    
    \quad \textbf{1.1:} Obtain the preliminary nuisance estimators $\widetilde \balpha$ and $\widetilde \bgamma$ by \eqref{eq:initial alpha}.
    
    \quad \textbf{1.2:} Obtain the preliminary estimator $\widetilde \bbeta$ of $\bbeta_0$ by solving $\Esc(\bbeta, \widetilde\balpha, \widetilde\bgamma ) = \bzero$ defined by \eqref{aeq}.

    \quad \textbf{1.3:} Obtain the calibrated nuisance estimators $\widehat \balpha_{\beta_j}$ and $\widehat \bgamma_{\beta_j}$ by \eqref{equ:cal:beta:alpha} using $\widetilde \balpha$, $\widetilde \bgamma$ and $\widetilde \bbeta$.

    \quad \textbf{1.4:} Obtain the DR estimator $\widehat \bbeta$ by solving $\Esc(\bbeta, \widehat\balpha_{\beta_j}, \widehat\bgamma_{\beta_j} ) = \bzero$ for $j=1,\ldots,q$.

       Constructing DR estimators $\TPR(c)$, $\FPR(c)$, $\ROC(u)$ and $\AUC$: 
  
      \quad \textbf{2.1:} Obtain the calibrated nuisance estimators $\widehat\balpha_{\roc,\widehat c_j}$ and $\widehat\bgamma_{\roc,\widehat c_j}$ by \eqref{equ:cal:roc}.

    \quad \textbf{2.2:} Obtain the DR estimators $\widehat \TPR(c)$ and $\widehat \FPR(c)$ of $\TPR(c)$ and $\FPR(c)$ by \eqref{equ:roc:dr:cal}.
    \textbf{Output:} 
    $\widehat \bbeta$, $\widehat \ROC(u) = \widehat \TPR\{\widehat \FPR^{-1}(u)\}$ for any $u\in[0,1]$, and $\widehat \AUC = \int_0^1 \widehat \ROC(u) {\rm d} u$. 
\end{algorithm}

\subsection{Calibrated estimation of the outcome parameters}\label{sec:method:cal:out}
 Inspired by  \cite{tan2020model} and \cite{dukes2020inference}, we construct calibrated nuisance estimators according to certain moment conditions, to correct for their regularization bias. First, we obtain preliminary estimators for $\bgamma$ in the imputation model $\E[Y\mid \X = \x] = g(\x\trans\bgamma)$ and $\balpha$ in density ratio $h(\x)=\exp(\x\trans\balpha)$ as
\begin{equation}
\begin{aligned}
\widetilde\balpha=&\argmin{\balpha\in\mathbb{R}^{q+p}}\Pschat_{\Sscsub\Cupsub\Tscsub}\{\rho_nS_i \exp (\X_i \trans \balpha) - \rho_N(1-S_i)\X_i \trans \balpha  \} + \lambda_{\alpha} \|\balpha\|_1;\\
\widetilde\bgamma=&\argmin{\bgamma\in\mathbb{R}^{q+p}}\Pschat\subSsc \{ -Y_i \X_i\trans \bgamma + G(\X_i\trans \bgamma)\}    + \lambda_{\gamma} \|\bgamma\|_1,
\end{aligned}    
\label{eq:initial alpha}
\end{equation}
where $G(a)=\int_{0}^ag(u) {\rm d} u$, $\rho_n=(N+n)/n$, $\rho_N=(N+n)/N$ and $\lambda_{\gamma}$, $\lambda_{\alpha}$ are two tuning parameters. Then we solve
$
\Esc(\bbeta, \widetilde\balpha, \widetilde\bgamma ) = \bzero 
$
to obtain a preliminary estimator $\widetilde\bbeta$. According to our discussion in Section \ref{sec:method:dr:cons}, $\widetilde\bbeta$ is consistent for the target $\bbeta_0$ when either the density ratio or the imputation model is correctly specified, while it may not achieve the desirable parametric rate $n^{-1/2}$ that is crucial for asymptotic inference. 

Denote the information matrix as $\bar\bSigma_{\bbeta} = \E\subSsc[\dot{g}(\A\trans \bbeta) \A \A\trans]$ and let $\widehat{\bSigma}_{\bbeta} = \Pschat\subTsc \dot{g}(\A_i\trans \bbeta) \A_i \A_i\trans$ where $\dot g(\cdot)$ is the derivative of $g(\cdot)$. Given $\e_j \in \mathbb{R}^q$  the $j$th unit vector in $\R^q$ with its $j$th element being one and other elements being zero, we now focus on estimating $\e_j\trans\bbeta_0$. Through asymptotic expansion, we find that the first order bias of $\sqrt{n} (\e_j\trans \widehat{\bbeta}- \e_j\trans \bbeta_0)$ can be potentially controlled by the two terms\footnote{We subscript the moment conditions and estimators with the symbol $\beta$ to distinguish them from those used for ROC estimation in Section \ref{sec:cal:roc}.}
\begin{equation}
\begin{split}
\Msc_{\beta_j,\alpha}(\balpha)&=: \big\|\Pschat_{\Sscsub\Cupsub\Tscsub}\widehat w_{ji}(\widetilde\bbeta)\dot{g}( \X_i\trans\widetilde\bgamma )\left\{\rho_n S_i \exp(\X_i\trans\balpha)\X_i-(1-S_i)\rho_N \X_i \right\}\big\|_{\infty} \,;\\
\Msc_{\beta_j,\gamma}(\bgamma)&=:\big\|\Pschat\subSsc\widehat w_{ji}(\widetilde\bbeta)\exp(\X_i \trans\widetilde\balpha)\{  Y_i - g(\X_i\trans\bgamma)\} \X_i \big\|_{\infty} \,,
\end{split}
\label{equ:mom:cons}    
\end{equation}
where $\widehat w_{ji}(\bbeta)=\e_j\trans \widehat\bSigma_{\bbeta}^{-1} \A_i$ and for simplicity, we denote by $\widehat w_{ji}=\widehat w_{ji}(\widetilde\bbeta)$. The detailed derivation is presented in Supplementary S3.3. It implies that when at least one nuisance model is correctly specified, one can reduce the impact of fitting bias of the nuisance estimators plugged in \eqref{aeq} through controlling \eqref{equ:mom:cons}. To this end, we conduct the following regularized regressions to calibrate the nuisance estimators:
\begin{equation}
\begin{split}
\widehat\bdelta_{\beta_j,\alpha}&=\argmin{\bdelta\in\mathbb{R}^{q+p}}\Pschat_{\Sscsub\Cupsub\Tscsub}\widehat w_{ji}\dot{g}( \X_i\trans\widetilde\bgamma )\Fsc_i(\bdelta;\widetilde\balpha) + \lambda_{\beta_j,\alpha} \|\bdelta\|_1 \, ;\\
\widehat\bdelta_{\beta_j,\gamma}&=\argmin{\bdelta\in\mathbb{R}^{q+p}}\Pschat\subSsc\widehat w_{ji}\exp(\X_i \trans\widetilde\balpha)\Gsc_i(\bdelta;\widetilde\bgamma)+ \lambda_{\beta_j,\gamma} \|\bdelta\|_1 \,,
\end{split}
\label{equ:cal:beta:alpha}
\end{equation}
and set $\widehat\balpha_{\beta_j} = \widetilde\balpha+ \widehat\bdelta_{\beta_j,\alpha}$, $\widehat\bgamma_{\beta_j} = \widetilde\bgamma + \widehat\bdelta_{\beta_j,\gamma}$, where
\begin{align*}
\Fsc_i(\bdelta;\balpha)& =  \rho_nS_i \exp \{\X_i \trans (\balpha+\bdelta)\} - \rho_N(1-S_i)\X_i \trans (\balpha+\bdelta) \, ;\\
\Gsc_i(\bdelta;\bgamma)&=-Y_i \X_i\trans (\bgamma+\bdelta) + G\{\X_i\trans (\bgamma+\bdelta)\}\,.
\end{align*}
Tuning strategy of the parameters $\lambda_{\beta_j,\alpha}$ and $\lambda_{\beta_j,\gamma}$ is in Supplementary S1 
and their rates are given in Section \ref{sec:thm}. Karush--Kuhn--Tucker (KKT) conditions of the LASSO problems \eqref{equ:cal:beta:alpha} guarantee the $\ell_{\infty}$-terms defined in (\ref{equ:mom:cons}) to be controlled by the parameters $\lambda_{\beta_j,\alpha}$ and $\lambda_{\beta_j,\gamma}$, i.e. $\Msc_{\beta_j,\alpha}(\widehat\balpha_{\beta_j}) \leq  \lambda_{\beta_j,\alpha}$ and $\Msc_{\beta_j,\gamma}(\widehat\bgamma_{\beta_j}) \leq  \lambda_{\beta_j,\gamma}$. Then we plug $\widehat\bgamma_{\beta_j}$ and $\widehat\balpha_{\beta_j}$ in \eqref{aeq} to solve $\widehat\bbeta$ from
$
\Esc(\bbeta, \widehat\balpha_{\beta_j}, \widehat\bgamma_{\beta_j}) = \bzero
$,
and obtain $\e_j\trans\widehat\bbeta$ as the estimator for $\e_j\trans\bbeta_0$. To obtain the estimator for the whole $\bbeta_0$, one can follow this procedure for $j = 1, \ldots, q$ to estimate each entry of $\bbeta_0$ separately and concatenate them together. With a little abuse of notation, we still denote the resulted vector estimator as $\widehat\bbeta$.

\begin{remark}
Different from \cite{tan2020model}, the weights $\widehat w_{ji}$'s may not be positive definite since we aim at transferring a regression model but not simple average parameters. For example, if $\A$ has entrywise symmetric distribution (except the intercept), it is possible that approximately half of $\widehat w_{ji}$'s are positive and the other half are negative. Hence, the loss function in  (\ref{equ:cal:beta:alpha}) can be irregular and ill-posed. To handle this issue, we follow a similar idea as \cite{liu2020doubly} to divide the samples into two sets with positive and negative only $\widehat w_{ji}$'s respectively and solve 
\begin{equation}
\begin{split}
\widehat\bdelta_{\beta_j,\alpha}^{s}&=\argmin{\bdelta\in\mathbb{R}^{q+p}}\Pschat_{\Sscsub\Cupsub\Tscsub}I_{ji}^s |\widehat w_{ji}| \dot{g}( \X_i\trans\widetilde\bgamma )\Fsc_i(\bdelta;\widetilde\balpha) + \lambda_{\beta,\alpha}^s \|\bdelta\|_1  \, ;\\
\widehat\bdelta_{\beta_j,\gamma}^s &= \argmin{\bdelta\in\mathbb{R}^{q+p}} \Pschat\subSsc I_{ji}^s |\widehat w_{ji}| \exp(\X_i \trans\widetilde\balpha)\Gsc_i(\bdelta;\widetilde\bgamma)+ \lambda_{\beta,\gamma}^s \|\bdelta\|_1 \\
\end{split}   
\label{equ:cal:sign}
\end{equation}
for $s = +, -$ separately instead of \eqref{equ:cal:beta:alpha}, where $I_{ji}^+ = \I(\widehat w_{ji}>0)$, $I_{ji}^- = \I(\widehat w_{ji} \leq 0)$, and $\lambda_{\beta_j,\alpha}^s,\lambda_{\beta_j,\gamma}^s$ are tuning parameters. Correspondingly, we take $\widehat\balpha^s_{\beta_j} = \widetilde \balpha+\widehat\bdelta_{\beta_j,\alpha}^s$ and  $\widehat\bgamma^s_{\beta_j}= \widetilde\bgamma+\widehat\bdelta_{\beta_j,\gamma}^s$, and solve $\e_j\trans\widehat\bbeta$ from
\[
\Pschat\subSsc \widehat h_{\beta_j}(\X_i) \A_i \{  Y_i -\widehat r_{\beta_j}(\X_i)\} + \Pschat\subTsc \A_i \{ \widehat r_{\beta_j}(\X_i)- g(\A_i\trans \bbeta) \} = \bzero,
\]
where $\widehat h_{\beta_j}(\X_i)=\I(\widehat w_{ji} > 0)\exp (\X_i \trans \widehat \balpha^+_{\beta_j} ) + \I(\widehat w_{ji} \leq0)\exp ( \X_i \trans \widehat \balpha^-_{\beta_j} )$ and $\widehat r_{\beta_j}(\X_i)=\I(\widehat w_{ji} > 0)g(\X_i\trans \widehat \bgamma ^+_{\beta_j}) +  \I(\widehat w_{ji} \leq 0)g(\X_i\trans \widehat \bgamma^-_{\beta_j})$, for $i = 1, \ldots, N+n$.  
\end{remark}

\begin{remark}
Different from \cite{tan2020model} and \cite{dukes2020inference}, we use an estimating procedure similar to Trans-Lasso proposed by \citep{li2020transfer} to construct the nuisance model estimators for both the model coefficients and ROC parameters. As is shown in our asymptotic and finite studies, this strategy can ensure more precise and stable nuisance estimators when the calibrated estimating equations have reduced sample size or effective sample size due to the sample splitting that appears in equations (\ref{equ:cal:sign}) and (\ref{equ:cal:roc}). 
\label{rem:1}
\end{remark}

\subsection{Calibrated estimation of the ROC parameters}\label{sec:cal:roc}

Based on the estimator $\widehat \bbeta$, we use \eqref{equ:roc:dr0} and \eqref{equ:roc:dr} to construct the doubly robust estimators for $\TPR(c)$ and $\FPR(c)$ simultaneously for all $c\in\mathbb{R}$. 
Similar to Section \ref{sec:method:cal:out}, { our goal is to derive calibrated nuisance estimators to mitigate the first order biases of $\sqrt{n} \{ \widehat \TPR(c) - \TPR(c)  \}$ and $\sqrt{n} \{ \widehat \FPR(c) - \FPR(c)  \}$. By the asymptotic analysis in Supplementary S3.4, 
with the initial estimators $\widetilde\balpha$ and $\widetilde\bgamma$, we find these biases can be controlled by constructing nuisance estimators to control the error terms} 
\begin{equation}
\begin{split}
&\Msc_{\roc,\alpha}(\balpha;c)=:  \big\|\Pschat_{\Sscsub\Cupsub\Tscsub}\I(\A_i\trans \widehat \bbeta \geq c)\dot{g}( \X_i\trans \widetilde\bgamma) \{\rho_n S_i  \exp(\X_i\trans\balpha)\X_i- \rho_N (1-S_i) \X_i \}\big\|_{\infty} \,;\\
&\Msc_{\roc,\gamma}(\bgamma;c)=:\big\|\Pschat\subSsc\I(\A_i\trans \widehat \bbeta \geq c)\exp(\X_i \trans\widetilde\balpha)\X_i\{  Y_i - g(\X_i\trans\bgamma)\}\big\|_{\infty} \,.
\end{split}
\label{equ:mom:cons:roc}    
\end{equation}
Controlling the terms in \eqref{equ:mom:cons:roc} is more challenging than \eqref{equ:mom:cons} since we need to control the empirical processes $\Msc_{\roc,\alpha}(\balpha;c)$ and $\Msc_{\roc,\gamma}(\bgamma;c)$ uniformly for $c\in\mathbb{R}$. 
The effective sample size for estimating TPR and FPR is also much smaller than $n$ for larger $c$, leading to additional statistical challenges. We propose a ``nearest quantile" strategy that only calibrates for a sequence of 
cutoff values (quantiles) decided from the data. 
Then each point on the ROC curve will be close enough to its nearest calibrated quantile. Hence the approximation bias between them could be negligible, and one can plug in the calibrated nuisance estimators of its nearest quantile to derive the doubly robust estimator for each point. 

\begin{figure}[htb!]
\centering
\caption{\label{fig:roc_ill} The nearest quantile strategy. The quantiles $\widehat c_j$'s are set to divide the source sample equally such that $n_{j} - n_{j+1} = n_{\min}$ and $n_m = n_{\min}$. The nuisance models are calibrated at these quantiles for the doubly robust estimation of $\TPR$ and $\FPR$. For each $\widehat c_j$, the effective sample size of its calibrating equations is labeled as $n_j$. 
For any point $c\in\mathbb{R}$, the nuisance models calibrated in the nearest quantile $t(c)$ will be used to construct the doubly robust estimators of $\TPR(c)$ and $\FPR(c)$.
}
\includegraphics[width=0.65\linewidth]{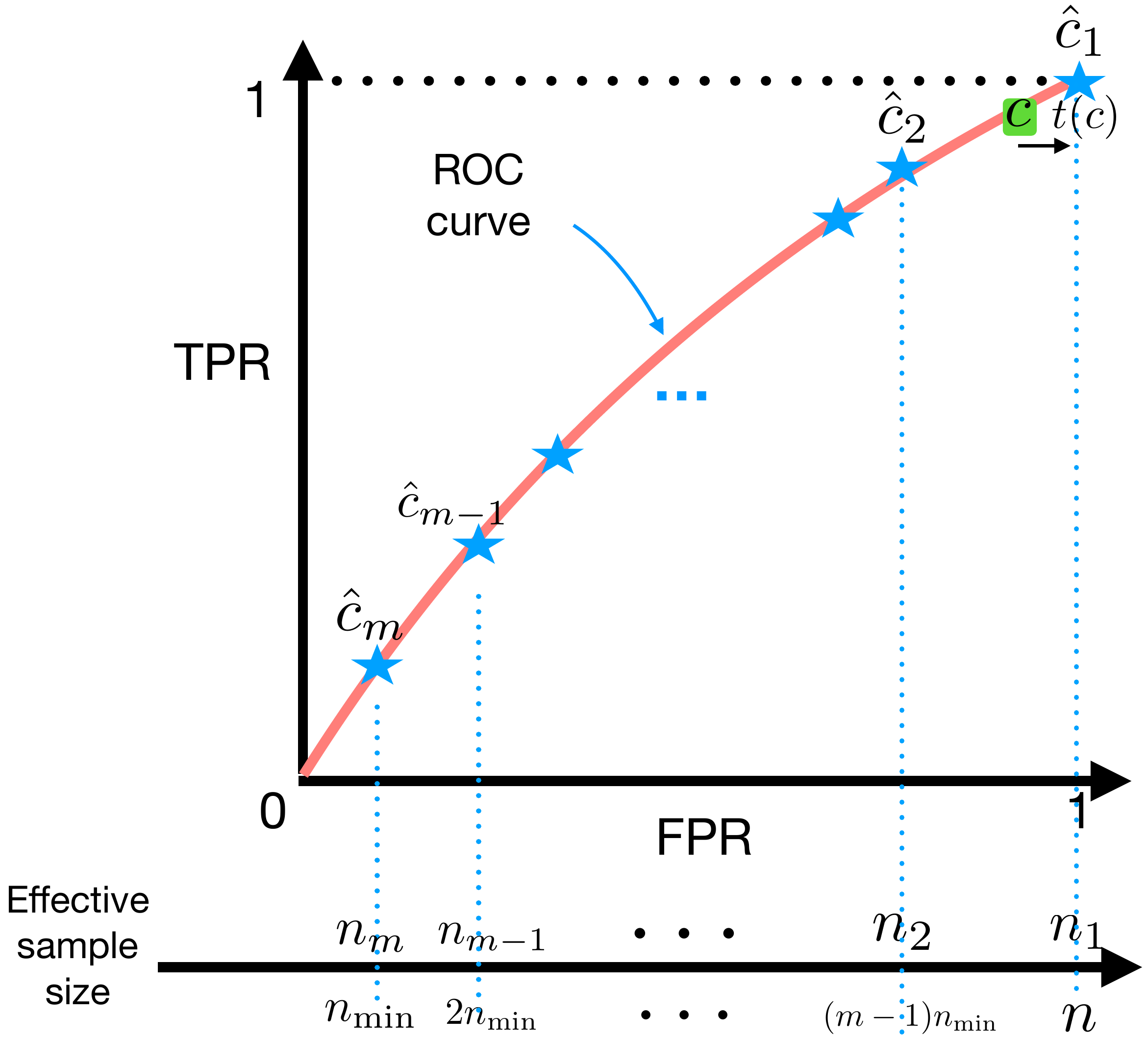} 
\end{figure}

Specifically, we split the source data into $m$ segments with equal sizes $n_{\min} = \lceil n/m \rceil$ according to the order of $\A_i\trans\widehat\bbeta$, where $ \lceil x \rceil$ is the smallest integer larger or equal to $x$. Without loss of generality, we assume that $\A_1\trans \widehat \bbeta \geq \A_2\trans \widehat \bbeta \geq \ldots \A_n\trans \widehat \bbeta$. We then choose the 
quantiles as $\widehat c_j = \A_{n_j}\trans\widehat\bbeta$ for $j \in [m]$ where $n_1 = n, n_2 = (m-1) n_{\min}$, $\ldots$, $n_m=n_{\min}$ as illustrated in Figure \ref{fig:roc_ill}. 
To control the moment terms \eqref{equ:mom:cons:roc} corresponding to each $\widehat c_j$, we propose to solve the following calibration equations for $c \in \Csc\subcal := \{\widehat{c}_j:j=1,\ldots,m\}$
\begin{equation}
\begin{split}
\widehat\bdelta_{\roc,\alpha, c}&=\argmin{\bdelta\in\mathbb{R}^{q+p}}\Pschat_{\Sscsub\Cupsub\Tscsub}\I(\A_i\trans \widehat \bbeta \geq c) \dot{g}( \X_i\trans\widetilde\bgamma )\Fsc_i(\bdelta;\widetilde\balpha)  + \lambda_{\roc,\alpha,c} \|\bdelta\|_1,\\
\widehat\bdelta_{\roc,\gamma,c}&=\argmin{\bdelta\in\mathbb{R}^{q+p}}\Pschat_\Sscsub\I(\A_i\trans \widehat \bbeta \geq c) \exp(\X_i \trans\widetilde\balpha)\Gsc_i(\bdelta;\widetilde\bgamma)+ \lambda_{\roc,\gamma,c} \|\bdelta\|_1,
\end{split}
\label{equ:cal:roc}
\end{equation}
and set $\widehat\balpha_{\roc,c}=\widetilde\balpha+\widehat\bdelta_{\roc,\alpha,t(c)}$,  $\widehat\bgamma_{\roc,c}=\widetilde\bgamma+\widehat\bdelta_{\roc,\gamma,t(c)}$ for any $c\in\mathbb{R}$ where  $t(c)=\argmin{c'\in\Csc\subcal}|c'-c|$. Here $\lambda_{\roc,\alpha,\widehat c_j}$ and $\lambda_{\roc,\gamma,\widehat c_j}$ are two series of $\ell_1$-penalty parameters to be tuned using the procedure described in Supplementary S1. 
We then construct the doubly robust estimators for $\TPR(\cdot)$ and $\FPR(\cdot)$ based on the calibrated nuisance estimators and plug the estimators $\widehat\balpha_{\roc,t(c)}$ and  $\widehat\bgamma_{\roc,t(c)}$
into \eqref{equ:roc:dr} to obtain 
\begin{equation}
\begin{aligned}
& \widehat \TPR(c) = \widehat \TPR(c; \widehat \bbeta, \widehat\balpha_{\roc,t(c)}, \widehat\bgamma_{\roc,t(c)},\widehat\balpha_{\roc,\widehat c_1},\widehat\bgamma_{\roc,\widehat c_1}),\\
& \widehat \FPR(c) = \widehat \FPR(c; \widehat \bbeta, \widehat\balpha_{\roc,t(c)}, \widehat\bgamma_{\roc,t(c)},\widehat\balpha_{\roc,\widehat c_1},\widehat\bgamma_{\roc,\widehat c_1}) \,.
\end{aligned}    
\label{equ:roc:dr:cal}
\end{equation}

\begin{remark}
The choice of $n_{\min}$ is critical. Large $n_{\min}$ can lead to non-negligible approximation error from $t(c)-c$, and hence imprecise calibration. Small $n_{\min}$ can result in overly small effective sample sizes for some calibrated estimating equations, says $\widehat c_j=\widehat c_m$ in equation \eqref{equ:cal:roc}, leading to high error rate in $\widehat\balpha_{\roc,\widehat c_j}$ and $\widehat\bgamma_{\roc,\widehat c_j}$. We study in Section \ref{sec:thm} this trade-off and derive an optimal $n_{\min}$ of order $n\suphalf(\log~np)^{1/3}$,  which can effectively remove the excessive first order bias in the nuisance estimators simultaneously for all $c$. 
\end{remark}
Finally, one can evaluate $\widehat\ROC(u)=\widehat\TPR\{{\widehat\FPR}^{-1}(u)\}$, and $\widehat\AUC=\int_0^1\widehat\ROC(u) {\rm d}u$, to estimate the ROC function and AUC. An illustration of this ``nearest quantile" strategy is presented in Figure \ref{fig:roc_ill}.

\section{Theoretical justification}\label{sec:thm}
Let $\|\u\|_0$ represent the number of non-zero entries of the input vector $\u$ and $\|\u\|_q$ be the $l_q$ norm of $\u$ for any $q>0$. We use $a_n = o(b_n)$ if $\lim_{n \rightarrow \infty} a_n/b_n = 0$, $a_n = O(b_n)$ if $\lim_{n \rightarrow \infty} a_n/b_n \leq C$ for some constant $C$, and $a_n \asymp b_n$ if  $a_n = O(b_n)$ and $b_n = O(a_n)$. Throughout, we denote by $\rho= \lim_{N,n \rightarrow \infty}(n/N)\suphalf$ and assume $\rho=O(1)$. We denote the population parameters by 
\begin{equation*}
\begin{aligned}
\bar\balpha  & = \argmin{\balpha\in\mathbb{R}^{q+p}} \big \{  \rho\subSsc \E\subSsc [ \exp (\X_i \trans \balpha)] -  \rho\subTsc \E\subTsc[ \X_i \trans \balpha  ] \big \} , \quad \bar\bgamma  =   \argmin{\bgamma\in\mathbb{R}^{q+p}} 
\E\subSsc [ -Y_i \X_i\trans \bgamma + G(\X_i\trans \bgamma)],
\end{aligned}    
\end{equation*}
where $\rho\subSsc = 1+\rho^{-2}$ and $\rho\subTsc = 1+\rho^{2}$, and for $j = 1,\ldots,q$, 
\begin{equation*}
\begin{aligned}
  \bar\balpha_{\beta_j}  & = \bar\balpha + \argmin{\bdelta\in\mathbb{R}^{q+p}} \e_j \trans \bSigma_{\bbeta_0}^{-1} \big\{ \rho\subSsc \E\subSsc  [\A_i \Fsc_{\Sscsub i}(\bdelta; \bar \balpha, \bar \bgamma)
 ] - \rho\subTsc \E\subTsc  [ \A_i
\Fsc_{\Tscsub i}(\bdelta;\bar \balpha,\bar \bgamma)   ] \big \}  \,,\\
 \bar\bgamma_{\beta_j} & = \bar\bgamma+  \argmin{\bdelta\in\mathbb{R}^{q+p}}  \e_j \trans \bSigma_{\bbeta_0}^{-1} \E\subSsc [ \A_i \exp(\X_i \trans \bar \balpha) \Gsc_i(\bdelta;\bar \bgamma)] \,,
\end{aligned}    
\end{equation*}
where $\Fsc_{\Sscsub i}(\bdelta;\balpha,\bgamma) = \dot{g}( \X_i\trans \bgamma ) \exp \{\X_i \trans ( \balpha+\bdelta)\}$ and $\Fsc_{\Tscsub i}(\bdelta;\balpha,\bgamma) = \dot{g}( \X_i\trans  \bgamma )  \X_i \trans (\balpha+\bdelta)$. 
We define $c_j$ as $1 - (m-j+1)n_{\min}/n$ population quantile of $\A_i\trans\bbeta_0$, for $j \in [m]$.
For $c \in \mathbb{R}$, we define 
\begin{equation*}
\begin{split}
\bar\balpha_{\roc,c} & = \bar \balpha +  \argmin{\bdelta\in\mathbb{R}^{q+p}} \big\{ \rho\subSsc \E\subSsc  [\I(\A_i \trans \bbeta_0 \geq c) 
 \Fsc_{\Sscsub i}(\bdelta; \bar \balpha, \bar \bgamma)  ] - \rho\subTsc \E\subTsc  [ \I(\A_i \trans \bbeta_0 \geq c) 
  \Fsc_{\Tscsub i}(\bdelta; \bar \balpha, \bar \bgamma) ] \big \} \,, \\
\bar\bgamma_{\roc,c} &= \bar \bgamma + \argmin{\bdelta\in\mathbb{R}^{q+p}}\E_\Sscsub [ \I(\A_i \trans \bbeta_0 \geq c ) \exp(\X_i \trans\bar \balpha) \Gsc_i(\bdelta;\bar \bgamma) ]\,.
\end{split}
\end{equation*}
For simplicity, we focus on the case without the need to split the samples through (\ref{equ:cal:sign}). Our analysis can be naturally generalized to this setting.  We first introduce assumptions for our theoretical analysis and remark on their appropriateness.

\begin{assume}
There exists a constant $\kappa>0$ such that $\max_{i\in\{1,\ldots,n+N\}}\|\X_i\|_{\infty}<\kappa$.
$\A_i$ belongs to a compact set, and its probability density is continuously differentiable. The eigenvalues of $\E\subTsc[\X_i\X_i\trans]$ and  $\E\subSsc[\X_i\X_i\trans]$ stay away from $0$ and $\infty$.
$\bbeta$ belongs to a compact parameter space $\Bsc$.
\label{asu:1}
\end{assume}

\begin{assume}
At least one of the following two conditions hold: (a) there exists $\balpha_0$ such that $h_0(\x) = p_{\X|S=0}(\x)/p_{\X|S=1}(\x)=\exp(\x\trans\balpha_0)$; or (b) there exists $\bgamma_0$ such that $r_0(\x)=\E\subSsc[Y|\X=\x]=\E\subTsc[Y|\X=\x]=g(\x\trans\bgamma_0)$.
\label{asu:model}
\end{assume}
\begin{assume}
There exists a positive constant $c_g$ such that 
$\min_{i\in\{1,\dots, n+N\}} \dot{g}(\X_i\trans\bar \bgamma) \geq c_g$.
{There exists a positive constant $C_{\delta}$ such that 
$\|\bar\bdelta_{\beta_j, \alpha}\|_1$, $\|\bar\bdelta_{\beta_j, \gamma}\|_1$, 
$\|\bar\bdelta_{\roc, \alpha, c}\|_1$, $\|\bar\bdelta_{\roc, \gamma, c}\|_1$, and $\|\bar\balpha\|_1$ are bounded by $C_{\delta}/\kappa$ for $j\in\{1, \dots, q\}$ and any $c$.}
\label{asu:3}
\end{assume}

\begin{assume}
Assume that $\|\bar\balpha\|_0,\|\bar\bgamma\|_0\leq s$ for some $s=o(n\suphalf/\log p)$, and $\bar \bdelta_{\beta_j, \alpha}$ and $\bar \bdelta_{\beta_j, \gamma}$ are $l_r$ sparse, i.e., $$\|\bar\bdelta_{\beta_j, \alpha}\|_r\leq R_r \quad \mbox{and}\quad \|\bar\bdelta_{\beta_j, \gamma}\|_r\leq R_r$$ 
for some $r\in [0,1]$ and $R_r>0$ satisfying $R_r( \log p/n)^{\frac{1-r}{2}} = o\{(\log p)^{-1/2}\}$. 
\label{asu:4}
\end{assume}

\begin{assume}
Assume that $s = o(n^{1/4}/\log p)$, $n_{\min} \asymp n\suphalf(\log~np)^{1/3}$,
and $\bar \bdelta_{\roc, \alpha, c}, \bar \bdelta_{\roc, \gamma, c}$ are $l_r$ sparse for all $c$, i.e., 
$\|\bar \bdelta_{\roc, \alpha, c}\|_r \leq R_{\roc, r}$ and $\|\bar \bdelta_{\roc, \gamma, c}\|_r \leq R_{\roc, r}$, 
where $r\in [0,1]$ and 
$$R_{\roc, r}  = o \left\{ \frac{n^{1/4 - r/4} }{(\log ~np)^{2/3- r/3}}\right\}.$$
\label{asu:5}
\end{assume}
\begin{remark}
Assumption \ref{asu:1} could be relaxed to sub-gaussian $\X_i$  by imposing on some moment regularity conditions. 
The assumption for $\A_i$ and $\A_i\trans\bbeta_0$ are commonly founded in handling the empirical process of the $M$-estimators and their model accuracy parameters \citep{gronsbell2018semi,gronsbell2020efficient}.
The assumption for eigenvalues of the covariance matrix is standard in high-dimensional settings \citep{buhlmann2011statistics,negahban2012unified}.
Assumption \ref{asu:model} is standard in the literature of doubly robust inference \citep[e.g.]{bang2005doubly}. Technical conditions in Assumption \ref{asu:3} have been commonly used for analyzing $l_1$-regularized GLM \citep[e.g.]{tan2020model}. 
It is reasonable in that when $r=1$ in Assumptions \ref{asu:4} and \ref{asu:5}, terms like $|\X_i\trans \bar\bdelta_{\beta_j, \alpha}|$ will be smaller than $\|\X_i\|_{\infty}\|\bar\bdelta_{\beta_j, \alpha}\|_1$, which is bounded under Assumptions \ref{asu:1}, \ref{asu:4}, and \ref{asu:5}.
\end{remark}

\begin{remark}

Assumptions \ref{asu:4} and \ref{asu:5} is on the approximate sparsity of 
$\bar\bdelta_{\beta_j, \alpha}$, $\bar\bdelta_{\beta_j, \gamma, c}$, $\bar\bdelta_{\roc, \alpha,c}$ and  $\bar\bdelta_{\roc, \gamma,c}$. The minimum effective sample size of the calibrating equations, $n_{\min}$ is picked to minimize the excessive bias rate through trading-off the estimation errors of the nuisance models and the approximation bias introduced by the samples in $\{i:\I(\A_i\trans \widehat \bbeta\geq \widehat c_m)\neq\I(\A_i\trans \widehat \bbeta\geq c)\}$ for any $c>\widehat c_m$; see Section \ref{sec:cal:roc} and Figure \ref{fig:roc_ill}. Such optimal choice ensures the sharpness of our sparsity assumptions.

\end{remark}

\begin{remark}

Different from \cite{tan2020model} and \cite{dukes2020inference}, we only impose $l_0$ (exact) sparsity assumptions on the preliminary estimators $\bar\balpha$ and $\bar\bgamma$ while imposing approximate sparsity assumptions on the increments of the calibrated nuisance parameters $\bar \bdelta_{\beta_j, \alpha}$, $\bar \bdelta_{\beta_j, \gamma}$, $\bar\bdelta_{\roc, \alpha, c}$ and $\bar\bdelta_{\roc, \gamma, c}$. These assumptions could be more reasonable than directly assuming the exact sparsity of the calibrated nuisance parameters since the calibrated estimating equations involve some additional weights on the GLM. Note that when the density ratio model is correctly specified, we have $\bar\bdelta_{\beta_j, \alpha} = \bar\bdelta_{\roc, \alpha, c} = \bzero$, and when the imputation model is correctly specified, $\bar\bdelta_{\beta_j, \gamma} = \bar\bdelta_{\roc, \gamma, c} = \bzero$. 
\end{remark}

Based on the above introduced assumptions, we present the main conclusions of our asymptotic analysis below. %
Theorems \ref{thm:1} and \ref{thm:2} give the $n\suphalf$-convergence and asymptotic expansion of our model coefficient estimator $\widehat\bbeta$ and model accuracy estimator $\widehat\ROC(u)$.

\begin{theorem}
Under Assumptions \ref{asu:1}--\ref{asu:4}, 
with the tuning parameters  $\lambda_{\alpha}$, $\lambda_{\gamma}$, $\lambda_{\beta_j,\alpha}$, and $\lambda_{\beta_j,\gamma}$ of order $(\log p/n)\suphalf$, for $j = 1, \ldots,q$, $\widehat\bbeta$ converges to $\bbeta_0$ in probability, and 
\[
\sqrt{n}\e_j\trans(\widehat\bbeta-\bbeta_0) = \frac{1}{\sqrt{n}}\sum_{i=1}^n  \e_j \trans \Wsc_{\Sscsub,\beta_j}(\D_i)+\frac{\rho}{\sqrt{N}}\sum_{i=n+1}^{n+N}  \e_j \trans \Wsc_{\Tscsub,\beta_j}(\D_i)+o_p(1),
\]
 converges weakly to $N(0,\sigma_j^2)$ for $j = 1, \ldots, q$, where
\begin{align*}
 \Wsc_{\Sscsub, \beta_j}(\D_i)&= \bar\bSigma_{\bbeta_0}^{-1} \A_i \exp(\X_i\trans \bar\balpha_{\beta_j}) \{  Y_i - g(\X_i\trans \bar\bgamma_{\beta_j})\}, \ 
 \Wsc_{\Tscsub, \beta_j}(\D_i) = \bar\bSigma_{\bbeta_0}^{-1} \A_i \{g(\X_i\trans \bar\bgamma_{\beta_j}) - g(\A_i\trans \bbeta_0) \} \,,
\end{align*}
and $\sigma_j^2 = {\rm Var}\big(\e_j \trans \Wsc_{\Sscsub,\beta_j}(\D_i)\big) + \rho^2 {\rm Var}\big(\e_j \trans \Wsc_{\Tscsub,\beta_j}(\D_i)\big)$.
\label{thm:1}
\end{theorem}

\begin{theorem}
Under Assumptions \ref{asu:1}--\ref{asu:5}, and with the tuning parameters  $\lambda_{\alpha}$, $\lambda_{\gamma}$, $\lambda_{\beta,\alpha}$, and $\lambda_{\beta,\gamma}$  of order $(\log p/n)\suphalf$ as well as $\lambda_{\roc,\alpha,\widehat c_j}$ and $\lambda_{\roc,\gamma,\widehat c_j}$ of order $n_jn^{-1}(\log~np/n_j)\suphalf$ for $j=1,\ldots,m$, we have
$$
\sqrt{n}\{\widehat\ROC(u)-\ROC(u)\}=\frac{1}{\sqrt{n}}\sum_{i=1}^n\Wsc_{\Sscsub,\roc}(u,\D_i)+\frac{\rho}{\sqrt{N}}\sum_{i=n+1}^{n+N}\Wsc_{\Tscsub,\roc}(u,\D_i)+o_p(1)
$$
uniformly for all $u\in(0,1)$, where $c_{u} = {\FPR}^{-1}(u)$, $\mu = \P\subTsc(Y=1)$, and for $\Ksc = \Ssc, \Tsc$, 
\begin{equation}
\begin{split}
 \Wsc_{\Kscsub,\roc}(u,\D_i) = &    \mu^{-1} \Wsc_{\Kscsub,{\rm TPR}}(c_{u},\bbeta_0, \bar \btheta_{\roc,c_{u}},\D_i) - (1-\mu)^{-1} \ROC(u) \Wsc_{\Kscsub,{\rm FPR}}(c_{u},\bbeta_0, \bar \btheta_{\roc,c_{u}},\D_i) \\
 & +  \{ \mu^{-1} \dot{\TP}(c_u, \bbeta_0) - (1-\mu)^{-1} \ROC(u) \dot{\FP}(c_u, \bbeta_0) \} \trans\Wsc_{\Kscsub, \beta}(\D_i) \,,    
\end{split}    
\label{equ:term1:roc}
\end{equation}
 where $\bar\btheta_{\roc,c_{u}}= (\bar\balpha_{\roc,c_{u}}\trans ,\bar\bgamma_{\roc,c_{u}}\trans)\trans$, 
 \begin{equation*}
\begin{aligned}
& \Wsc_{\Kscsub,{\rm TPR}}(c,\bbeta_0, \bar \btheta_{\roc,c},\D_i) = \Wsc_{\Kscsub 1}(c,\bbeta, \bar \btheta_{\roc,c},\D_i) - \TPR(c,\bbeta_0) \Wsc_{\Kscsub 1}(-\infty,\bbeta, \bar \btheta_{\roc,c},\D_i) \,, \\
& \Wsc_{\Kscsub, {\rm FPR }}(c,\bbeta_0, \bar \btheta_{\roc,c},\D_i)   = \Wsc_{\Kscsub 2}(c,\bbeta_0, \bar \btheta_{\roc,c},\D_i) - \FPR(c,\bbeta_0) \Wsc_{\Kscsub 2}(-\infty,\bbeta_0, \bar \btheta_{\roc,c},\D_i) \,, \\
& \Wsc_{\Sscsub 1}(c,\bbeta, \btheta,D_i) = \I(\A_i\trans \bbeta \geq c) \exp(\X_i\trans  \balpha) \{Y_i-g(\X_i\trans \bgamma)\} - \E\subSsc\I(\A\trans  \bbeta \geq c) \exp(\X\trans \balpha) \{Y-g(\X\trans \bgamma)\} \,, \\
& \Wsc_{\Tscsub 1}(c,\bbeta, \btheta,D_i) = \I(\A_i\trans \bbeta \geq c) g(\X_i\trans \bgamma) - \E\subTsc\I(\A_i\trans \bbeta \geq c) g(\X_i\trans  \bgamma) \,, \\
& \Wsc_{\Sscsub 2}(c,\bbeta, \btheta,\D_i) = - \Wsc_{\Sscsub 1}(c,\bbeta, \btheta,\D_i) 
\,, \\
& \Wsc_{\Tscsub 2}(c,\bbeta, \btheta, \D_i) = \I(\A_i\trans \bbeta \geq c) \{1-g(\X_i\trans \bgamma)\} - \E\subTsc\I(\A_i\trans \bbeta \geq c) \{1-g(\X_i\trans \bgamma)\} \,, \\
& \dot{\TP}(c, \bbeta) = \partial \TP (c, \bbeta) / \partial \bbeta  =  \partial \P\subTsc(Y=1,\A\trans\bbeta \geq c) / \partial \bbeta\,, \\
& \dot{\FP}(c, \bbeta) = \partial \FP (c, \bbeta) / \partial \bbeta  =  \partial \P\subTsc(Y=0,\A\trans\bbeta \geq c) / \partial \bbeta\,.
\end{aligned}    
\end{equation*}
Consequently, $\{\sqrt{n}\{\widehat\ROC(u)-\ROC_0(u)\}:u\in(0,1)\}$ weakly converges to a Gaussian process with mean $0$, and $\sqrt{n}(\widehat{\AUC} - \AUC)$ weakly converges to a normal distribution with mean $0$. 
\label{thm:2}
\end{theorem}
Similar to the results of \cite{hahn1998role}, one can show that when both nuisance models are correct, our estimators of the model coefficients and the accuracy parameters attain semiparametric efficiency. While different from doubly robust estimators constructed with $n\suphalf$-consistent nuisance estimators under low dimensionality studied in previous literature \citep[e.g.]{bang2005doubly,shu2018improved}, the nuisance estimators in our case do not contribute non-negligible variability to $\widehat\bbeta$ and $\widehat\ROC(u)$ even when they are misspecified. This is due to the calibrating procedures introduced to remove the excessive bias caused by them. {As a benefit of this, we do not need to perturb or account for the variation of nuisance estimators in the bootstrap procedure as pointed out in Remark S.2 
of Supplementary S2, 
which is highly desirable given the high-dimensionality of $\bgamma$ and $\balpha$.}

When $N\gg n$, i.e., the size of target samples is much larger than the labeling size, influence of the target samples, i.e., ${\rho}N^{-1/2}\sum_{i=n+1}^{n+N}  \e_j \trans \Wsc_{\Tscsub,\beta_j}(\D_i)$ and ${\rho}N^{-1/2}\sum_{i=n+1}^{n+N}\Wsc_{\Tscsub,\roc}(u,\D_i)$ will be asymptotically negligible as $\rho\rightarrow 0$. So the asymptotic variance of our estimators can be reduced compared with the case with $N\asymp n$. Similar phenomenons could be found in the recent literature of semi-supervised inference \citep[e.g.]{kawakita2013semi,chakrabortty2018efficient}. Inspired by \cite{gronsbell2018semi}, when $g(\A\trans\bbeta)$ is correctly specified on the target population, i.e. $\E\subTsc[Y\mid\A]=g(\A\trans \bbeta_0)$, the coefficients of the score function $\Wsc_{\Kscsub, \beta_j}(\D_i)$ in (\ref{equ:term1:roc}) become zero. That means the variability of $\widehat\bbeta$ will not contribute to the variability of $\widehat\ROC(u)$. 

\section{Simulation studies}\label{sec:simu}

We conducted simulation studies to validate the proposed point and interval estimation procedures. We set $q=4$, $p=100$ or $200$, $N=3000$, and $n=600$ throughout. To generate $\X$ allowing for potential model mis-specifications in the density ratio and outcome models, we first generate $\U = (U_1,U_2, \ldots , U_{p+q})\trans$ with $U_1 = 1$ and  $U_2, \ldots, U_{p+1}$ generated from independent $N(0, 1)$ truncated to the interval $(2.5, 2.5)$ then standardized to zero mean and unit variance. We let $\X$ be functions of $\U$ and subsequently generate $Y$ and $S$ (indication of source samples) given $\X$ from Bernoulli distributions independently with the following three data-generating configurations: (i) both nuisance models are correctly specified; (ii) the density ratio model is correct while the imputation model is misspecified; and (iii) the imputation model is correct while the density ratio model is misspecified. The details are given in Supplementary S4.1.

We compare our DRAMATIC method with two existing methods: {(i) the importance weighting method (IW)  detailed in Algorithm S.1 
that adapts the strategy of \cite{huang2007correcting} to our setup; and (ii) the imputation only method (IM) detailed in Algorithm S.2 
presented in Supplementary S4.2 that is the dual method of IW and also considered by \cite{liu2020doubly} as a benchmark.} We set $n_{\min}=120$ for DRAMATIC, select the penalty parameters $\lambda_{\alpha}$ and $\lambda_{\gamma}$ using cross-validation from the range $[0.01(\log p/n)\suphalf,0.5(\log p/n)\suphalf]$, and select $\lambda_{\beta,\alpha}$, $\lambda_{\beta,\gamma}$, $\lambda_{\roc,\alpha,\widehat{c}_j}$ and $\lambda_{\roc,\gamma,\widehat{c}_j}$ using the tuning procedure introduced in Supplementary S1. 

For $p=100$, we report in the main text the average bias (Bias), root of average mean square error (rMSE), and average coverage probability (CP) of $95\%$ CI for the estimators based on $1000$ repetitions for each configuration. Results of estimating $\bbeta_0$ are presented in Table S1 of Supplementary S4.3 and those of AUC,  $\ROC(0.1)$ and $\ROC(0.2)$ are presented in Table~\ref{tab:auc}. {We also conducted simulation for $p=200$, which yielded a similar pattern to $p=100$ with results in Tables S2 and S3 presented in Supplementary S4.3.} Table S1 shows that DRAMATIC performs consistently well in estimating $\bbeta$. The interval estimators of DRAMATIC have proper coverage probabilities located in the $\pm0.05$ range of the nominal level. Under Configuration (ii) where the imputation model is misspecified, IM has large biases for some coefficients. For example, it has absolute bias $0.24$ for $\beta_3$, while DRAMATIC has biases smaller than $0.03$. Under Configuration (iii) where the density ratio model is misspecified, IW yields an unsatisfactory bias on $\beta_3$, which is more than $80\%$ of the MSE. On the contrary, the biases of DRAMATIC do not exceed $0.03$. Throughout the three configurations, the biases of DRAMATIC are smaller than $23\%$ of the rMSE for all coefficients $\beta_j$'s, while the performances of IM and IW are not robust, especially when the models are misspecified. Although they have low bias and rMSE under some cases, the bias of IM can be $95\%$ of the rMSE for $\beta_3$ in Configuration (ii) and of IW can be $90\%$ of the rMSE for $\beta_3$ in Configuration (iii). Even under Configuration (i) where both nuisance models are correct, DRAMATIC is still the most efficient method achieving the smallest rMSE for all $\beta_j$'s due to its debiasing correction. For example, DRAMATIC attains $169\%$ relative efficiency compared to IM on $\beta_1$ and $160\%$ relative efficiency compared to IW on $\beta_2$. 

From Table~\ref{tab:auc}, we see that DRAMATIC achieves the smallest bias and rMSE for almost all model accuracy parameters in the three configurations. In addition, under all the settings, our interval estimators have proper coverage probabilities located in $\pm0.02$. 
Under Configuration (i), DRAMATIC yields absolute biases no larger than $0.004$, while IM and IW can produce absolute bias larger than $0.02$, exceeding five times of DRAMATIC. Under Configurations (ii) and (iii), the performances of IM and IW are getting even worse, with absolute biases arriving at about $0.05$. Although the model of IM is correct under Configuration (ii) and the model of IW is correct under Configuration (iii), they are still not able to produce estimators with low bias due the excessive errors of the high-dimensional nuisance models. For example, under Configuration (ii), the bias of IM accounts for $68\%$ of the rMSE on AUC and $81\%$ of the rMSE on ROC($0.1$). Under Configuration (iii), the bias of IW accounts for $53\%$ of its rMSE on AUC. On the contrary, under all the configurations, DRAMATIC outputs a low bias of no more than $2\%$ of the magnitude of the true parameters and only occupy a small proportion of its rMSE. Consistent with our theoretical findings, these simulation results provide finite sample validation that  DRAMATIC is robust and efficient in model estimation and evaluation when at least one nuisance model is correctly specified; and substantially outperforms existing IM and IW estimators. 

\begin{table}[th] 
\centering 
\caption{\label{tab:auc} Average bias (Bias), root of average mean square error (rMSE), and average coverage probability (CP) of $95\%$ CI for the estimations of AUC and ROC with $p=100$, $n = 600$, and $N=3000$ under Configurations (i), (ii) and (iii) with $1000$ repetitions for each configuration.} 
\begin{tabular}{c l|cc|cc|ccc}
  \hline \hline 
\multirow{2}{*}{Configuration }  &  & \multicolumn{2}{c|}{IM} &  \multicolumn{2}{c|}{IW} &  \multicolumn{3}{c}{DRAMATIC}  \\  \cline{3-9}
 & & Bias & rMSE & Bias & rMSE & Bias & rMSE & CP \\
  \hline
\multirow{3}{*}{ (i)} & AUC &  -0.018 & 0.025 & \textbf{-0.003} & 0.033 & 0.004 & \textbf{0.024} & 0.932  \\  
& ROC(0.1) & 0.018 & 0.057 & 0.021 & 0.058 & \textbf{-0.004} & \textbf{0.036} & 0.971  \\  
& ROC(0.2) & -0.021 &  \textbf{0.035} & -0.018 & 0.037 & \textbf{0.001} & 0.043 & 0.954 \\  \hline
\multirow{3}{*}{ (ii)} & AUC &  -0.033 & 0.040 & -0.009 & 0.038 & \textbf{0.004} & \textbf{0.032} & 0.931  \\  
& ROC(0.1) & -0.054 & 0.060 & -0.050 & 0.059 & \textbf{0.008} & \textbf{0.056} & 0.947  \\  
& ROC(0.2) & -0.052 & 0.061 & -0.049 & 0.061 & \textbf{0.014} & \textbf{0.060} & 0.953 \\  \hline
\multirow{3}{*}{(iii)} & AUC & -0.029 & 0.034 & -0.024 & 0.033 & \textbf{0.000} & \textbf{0.019} & 0.925  \\  
& ROC(0.1) & -0.040 & 0.047 & -0.040 & 0.048 & \textbf{-0.007} & \textbf{0.039} & 0.938  \\  
& ROC(0.2) & -0.042 & 0.051 & -0.042 & 0.053 & \textbf{-0.001} & \textbf{0.039} & 0.946 \\  \hline
\end{tabular}
\end{table} 

\section{Application to electronic medical records studies}\label{sec:data analysis}

Type II diabetes (T2D), mainly caused by insulin resistance or insufficiency, is a common disease with an increasing prevalence over recent years. Existing genetic studies and genome-wide association studies (GWAS) have detected and identified hundreds of genetic variants associated with T2D, which enables the use of polygenic risk score (GRS) for T2D risk prediction \citep{he2021comparisons} and promises precision medicine approaches to preventing and managing the disease.

Using the Mass General Brigham (MGB) biobank data \citep{castro2022mass}, 
we aim to construct and evaluate a genetic risk prediction model of T2D combining GRS and demographic information. While genome-wide single nucleotide polymorphism (SNP) and demographic information are available on biobank patients, the T2D status is only available among a subset of $n=449$ patients whose medical records were manually reviewed in 2014. To create labels for multiple phenotypes efficiently and simultaneously, patients with diagnostic codes for several other diseases such as coronary artery disease were jointly sampled, leading to sampling not completely random. The target population consist of $N=2000$ subjects randomly sampled from the biobank participants in 2021 with their feature data updated in 2021. Both the sampling scheme and the different time window of data collection can cause shift of EHR features between the source and target samples. The data shift is particularly pronounced in this setting because around 2015, the EHR system at MGB was switched and the International Classification of Diseases (ICD) system was changed from version 9 to version 10. 

We aim at estimating and evaluating the logistic model of $Y\sim \A=(1,A_1,A_2,A_3)$ on the target population, where $A_1$ is the GRS of T2D, $A_2$ indicates gender being female, $A_3$ indicates race being white. To create the GRS of T2D, we utilize the GWAS results of \cite{mahajan2018fine} to choose SNPs significantly associated with T2D (with $p$-value below $10^{-10}$), prune the highly corrected SNPs (due to linkage disequilibrium), and then combine the selected 65 SNPs weighted by their beta coefficients reported in the paper. The adjustment features $\X$ are taken as the union of $\A$ and the health-utilization-adjusted (through partial linear regression) log-counts ($u\xrightarrow{}\log(u+1)$) of $p=143$ T2D-relevant diagnostic and procedure codes selected among the whole EHR features using the clinical knowledge extraction approach proposed by \cite{hong2021clinical}, independently from the source and target samples. 

To validate the transfer learning and evaluation approaches, we randomly sample and label $267$ subjects from the target population to construct a validation data set. Following \cite{liu2020doubly}, we fit $Y \sim \A$ using the validation data to obtain $\widehat\bbeta_{\rm val}$ as a ``gold standard" benchmark and use the following metrics to evaluate the performance of a transfer learning estimator $\widehat \bbeta$: (1) root of mean square prediction error (rMSPE) against $\widehat\bbeta_{\rm val}$ evaluated on the target data:
$\big[{\Pschat{\subTsc}\{g(\A_i\trans \widehat\bbeta_{\rm val} )- g(\A_i\trans \widehat\bbeta)\}^2}\big]\suphalf$; (2) classifier’s correlation (CC) with that of $\widehat\bbeta_{\rm val}$:
$
{\rm Corr}{\subTsc}\big\{ \I( g(\A_i\trans \widehat\bbeta_{\rm val}) \geq  \Pschat{\subTsc}  g(\A_{i'}\trans \widehat\bbeta_{\rm val} )),  \I( g(\A_i\trans \widehat \bbeta) \geq  \Pschat{\subTsc}  g(\A_{i'}\trans \widehat \bbeta))  \big\} \,,
$
where ${\rm Corr}{\subTsc}(\cdot,\cdot)$ represents the empirical Pearson correlation on the target population; and (3) false classification rate (FCR) of $\widehat \bbeta$'s classifier against that of $\widehat\bbeta_{\rm val}$:
$\Pschat{\subTsc}\big(  \I( g(\A_i\trans \widehat\bbeta_{\rm val}) \geq  \Pschat{\subTsc}  g(\A_{i'}\trans \widehat\bbeta_{\rm val}) )  \neq  \I( g(\A_i\trans \widehat \bbeta) \geq  \Pschat{\subTsc}  g(\A_{i'}\trans \widehat \bbeta))  \big)$. 
To evaluate the performance of estimating the ROC curve, we estimate the ROC and AUC of $\A\trans\bbeta$ on the validation data by cross-validation and use it as the ``gold standard". The total variation (TV) of any estimated ROC curve to this gold standard curve is calculated as the evaluation metric. Methods under comparison include our proposed DRAMATIC approach, the IW and IM approaches, 
and the naive source estimation (Source) obtained by fitting and evaluating the regression model simply using the labeled source samples.

Estimation of $\bbeta$ obtained using these included methods are presented in Table S4 of  Supplementary S4.3, and their performance evaluated against the validation estimator $\widehat\bbeta_{\rm val}$ (referred as ``Target") presented in Table S5. Among all methods under comparison, the DRAMATIC estimator achieves the closest estimated prevalence, the smallest rMSPE, the largest classifier’s correlation, and the smallest false classification rate evaluated against the validation estimator. For example, in terms of rMSPE, our estimator has a $68\%$ smaller error than Source, $54\%$ smaller than IW, and $34\%$ smaller than IM.

\begin{figure}[htb!]
\centering
\caption{\label{fig:roc} Estimated ROC curves. Source: naive source data estimator; DRAMATIC: our proposed doubly robust method; IW: importance weighting method; IM: imputation based method; Target: benchmark obtained using the validation samples.
}
\includegraphics[width=0.6\linewidth]{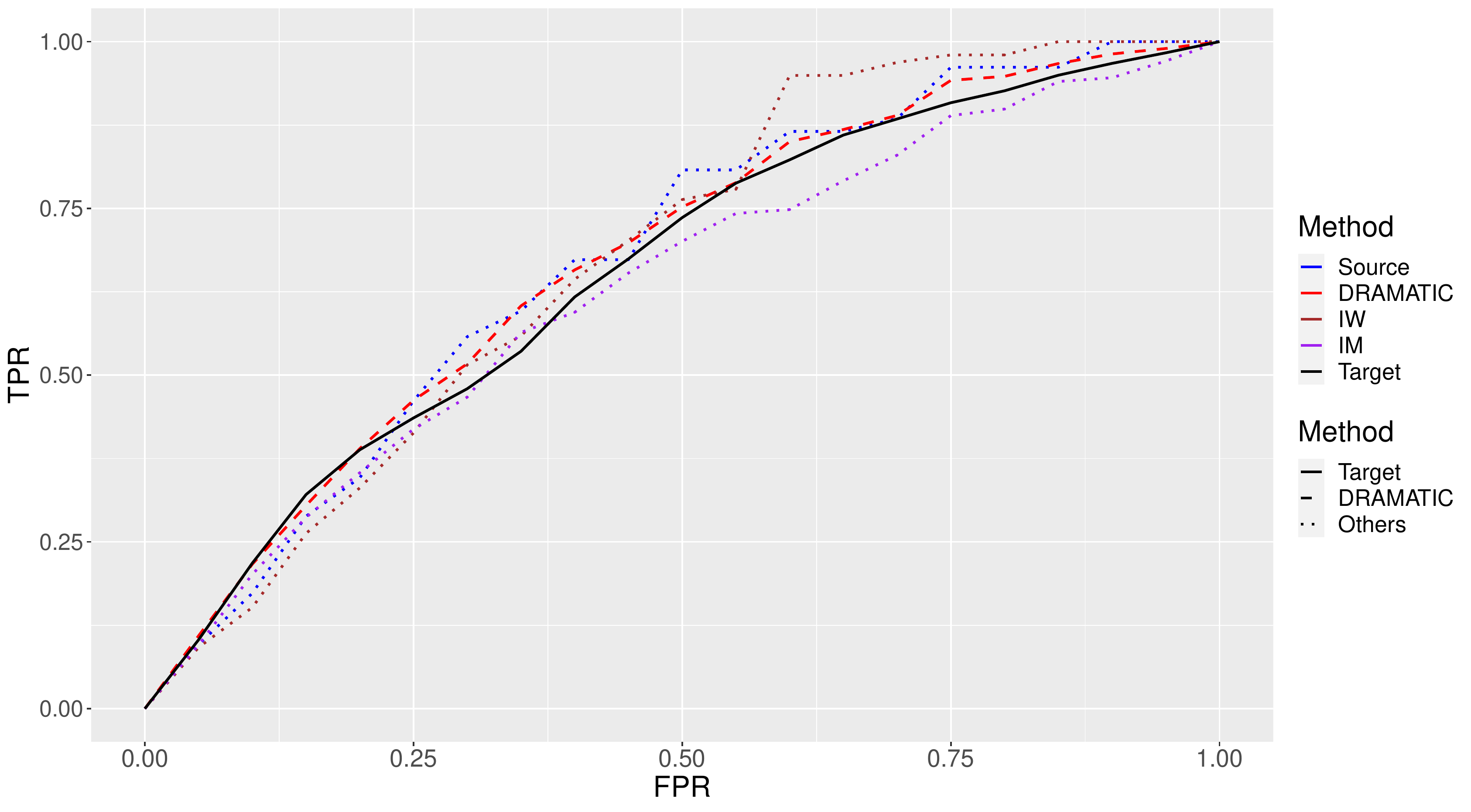} 
\end{figure}

We further plotted the estimated ROC curves in Figure \ref{fig:roc}. One can see that the ROC curve estimated using DRAMATIC (the red dashed line) is the closest one to the validation benchmark (the black solid line) among all methods. This is strictly verified by the total variation (TV) distance from the ``Target" ROC estimator presented in Table \ref{tab:real:roc}. In specific, the TV error of our approach is $41\%$ smaller than that of the Source estimator, $58\%$ smaller than IW, and $32\%$ smaller than IM. As shown in Table \ref{tab:real:roc}, our approach also produces the closest estimation in AUC, ROC$(0.1)$ and ROC$(0.2)$ to the Target than other transfer learning or source estimators. In addition, our 95\% CI estimators of AUC and ROC attain moderate and reasonable length and correctly cover the validation estimators. All these promising results illustrate that DRAMATIC is more robust and efficient than existing methods in transfer learning and model evaluation under the distributional shift of high-dimensional adjustment features. 

\begin{table}[htb!]
\footnotesize
\centering
\caption{\label{tab:real:roc} 
The results of estimating AUC, ROC$(0.1)$ and ROC$(0.2)$. 
Total variation distance from the validation estimator (TV from Target) of the ROC curve, and estimators of AUC, ROC$(0.1)$ and ROC$(0.2)$, attached with their 95\% CIs constructed using our method. 
}
\label{tab:roc}
\begin{tabular}{ccccccc}
\hline
\hline
          & Source & DRAMATIC (95\% CI) & IW & IM & Target \\ \hline
TV from Target &  0.032 & {\bf 0.019} & 0.045 & 0.028 & 0 \\
AUC &  0.677 & 0.671 (0.622, 0.720) & 0.673 & 0.626 & 0.656 \\
ROC$(0.1)$ &  0.173 & 0.217 (0.152, 0.281) & 0.152 & 0.201 & 0.218\\
ROC$(0.2)$ &  0.346 & 0.390 (0.310, 0.470) & 0.331 & 0.354 & 0.388\\
\hline\hline\\[-1.8ex] 
\end{tabular}
\end{table}

\section{Discussion}
In this paper, we propose DRAMATIC, a transfer
learning approach to estimate the coefficients and evaluate the ROC of the logistic model on a target population with no observations of $Y$ under the distributional shift of high-dimensional covariates. Our method is doubly robust in the sense that either a correctly specified density ratio model or a correct imputation model can lead to $n\suphalf$-consistent estimators under certain sparsity conditions. The main challenge of constructing a doubly robust estimator of the ROC curve under the high-dimensional nuisance model regime is to remove the excessive regularization bias for all points on the ROC curve simultaneously. We realize this through a novel ``nearest quantile" strategy that is statistically effective and computationally efficient, as introduced in Section \ref{sec:cal:roc}. In addition, unlike existing doubly robust estimators with high-dimensional nuisance models, our method does not require the calibrated nuisance parameters to be exactly sparse. Instead, we make assumptions on their $l_r$ norms for arbitrary $r\in [0,1]$, which is more general and reasonable than the $l_0$ sparsity.  

We note that other evaluation prediction and classification metrics, such as the overall misclassification
rate and the Brier score \citep[e.g.]{gronsbell2020efficient} can be further considered in our setting. Our high-dimensional calibrated method for curve estimator involving step functions could be potentially used in other settings such as the diagnose curve evaluation of the survival models. Besides, we target a classification model with the dimension of $\A$, $q$ being low and fixed. While there are also real-world interests in transferring evaluation of a high-dimensional sparse model where $q$ grows with $n$ and $N$. This is an open and challenging problem due to the excessive regularization error of $\bbeta_0$ in the high-dimensional case. 


\bibliographystyle{chicago}
\bibliography{library}

\end{document}